\documentclass[aps,twocolumn,prl,tightenlines,floatfix,showpacs]{revtex4}
\usepackage[dvips]{graphicx}
\usepackage[english]{babel}  
\usepackage{amsmath}  
\usepackage{amssymb}  
\usepackage{slashed}  
\usepackage{feynmf}
\usepackage{marginnote}

\begin{document}

\title{The Two Component Optical Conductivity in the Cuprates:
A Necessary Consequence of Preformed Pairs}

\author{Dan Wulin $^{1}$, Hao Guo $^{2}$,
Chih-Chun Chien$^{3}$ and K. Levin$^{1}$}

\affiliation{$^1$James Franck Institute and Department of Physics,
University of Chicago, Chicago, Illinois 60637, USA}

\affiliation{$^2$Department of Physics, University of Hong Kong,
Hong Kong, China}

\affiliation{$^3$Theoretical Division, Los Alamos National Laboratory, MS B213, Los Alamos, NM 87545, USA}

\date{\today}
\pacs{BHR1204}

\begin{abstract} 
We address how the finite frequency real conductivity $\sigma(\omega)$
in the underdoped
cuprates is affected by the pseudogap, contrasting the
behavior above and
below $T_c$. The f-sum rule is
analytically shown to hold. Here we presume the pseudogap
is associated with non-condensed pairs 
arising from stronger-than-BCS attraction.
This leads to both a Drude and a
mid infrared (MIR) peak, the latter 
associated with the energy needed to break
pairs. These 
general characteristics appear consistent with experiment.
Importantly, there is no more theoretical flexibility
(phenomenology) here than
in BCS theory; the
origin of the two component conductivity we find is robust. 
\end{abstract}

\maketitle

The behavior of the in-plane ac conductivity 
$\sigma(\omega) $
in the underdoped high temperature superconductors
has raised a number of puzzles \cite{TimuskRMP} for theoretical scenarios
surrounding the origin of the mysterious pseudogap.
At the same time,
there has been substantial recent progress in establishing
experimental constraints on  
the inter-play of the pseudogap and $\sigma(\omega)$\cite{Basov3} .
A key feature of 
$\sigma(\omega)$ is its two component nature 
consisting of a 
``coherent" Drude
like low $\omega$ feature followed by an approximately $T$-independent
mid-infrared (MIR) peak \cite{TimuskRMP,Basov3,Bontemps}.
The latter
``extends to the pseudogap boundary in the phase diagram
at $T^*$. Moreover a softening of the
MIR band with doping [scales with] the decrease in the pseudogap
temperature $T^*$" \cite{Basov3}.
Crucial to this picture is that ``high $T_c$ materials are in
the clean limit and that ... the MIR feature is seen above and
below $T_c$ "\cite{Tanner3}. Thus, it appears that this feature
is not associated with disordered superconductivity and
related momentum non-conserving processes,
but rather it is ``due to the unconventional nature of the 
[optical] response" \cite{TimuskRMP}.

It is the purpose of this paper to address these related observations in
the context of a 
preformed pair Gor'kov based theory that extends
BCS theory to the strong attraction limit \cite{CSTL05}.
Our expressions for $\sigma(\omega)$ are
equivalent to their BCS analogue when the pseudogap vanishes.
This approach is microscopically based and
the level of phenomenological flexibility \cite{OurBraggPRL,Ourconduct}
is no more than
that associated with transport in strict
BCS superconductors. 
Alternative mechanisms for the two component optical response include
Mott related physics \cite{Millis5} and d-density
wave \cite{Benfatto2} approaches, which have
acknowleged inconsistencies \cite{Benfatto3},
as well as approaches that build on inhomogeneity effects \cite{Orenstein}.
Distinguishing our approach 
is its very
direct association with the pseudogap. 
In an evidently less transparent way, a two component
response arises numerically \cite{Millis5} in
the presence of
Mott-Hubbard correlations above $T_c$.
However, experiments 
show how the MIR feature must persist
in the presence of superconductivity, suggesting that
pseudogap physics  affects superconductivity below $T_c$, as found
here.

Unique is our capability to
address both the normal (pseudogap) and superconducting phases.
Moreover,
we are also able to establish \cite{OurBraggPRL,Ourconduct}
compatibility with the
transverse f-sum rule without problematic negative conductivity
\cite{Millis5} contributions. 
Finally, our approach is to be distinguished from
the phase fluctuation scenario that appears
problematic in light of recent optical data
related to imaginary THz conductivity \cite{Bilbro}.
In experimental support of our scenario is the claim
based on $\sigma(\omega)$ data \cite{Deutscher2}
that the
``doping dependence suggests a smooth transition from a BCS mode
of condensation in the overdoped regime to a different mode in
underdoped samples, [as] in the
case of a BCS to Bose-Einstein crossover."

Our 
analysis leads to the following physical
picture: the presence of non-condensed pairs both above and
below $T_c$ yields an MIR peak. This peak occurs around the energy needed to break
pairs and thereby create conducting fermions. Its position 
is doping dependent, and only weakly temperature dependent, 
following the weak $T$ dependence of the excitation gap $\Delta(T)$.
The
relatively high frequency spectral weight from these
pseudogap effects, present in the normal
phase, is transferred to the condensate as $T$ decreases below $T_c$,
leading to
a narrowing of the low $\omega$ Drude feature,
as appears to be experimentally observed.
Even relatively poor samples are in the clean limit \cite{Tanner3,TimuskRMP}, 
so that
an alternative pair creation/annihilation contribution associated with
broken translational invariance 
cannot be invoked to explain the observed MIR absorption.

Before doing detailed calculations,
it is possible to anticipate the behavior of
$\sigma(\omega)$ at a physical level. In addition to the
$\omega \equiv 0$ condensate contribution,
the $\omega \neq 0$ conductivity consists of two terms, the more
standard one associated
with scattering of fermionic quasiparticles and the other associated with
the breaking of the pairs. 
The term associated with
the scattering of fermionic
quasi-particles gives rise to the usual Drude peak.
In the presence of stronger than BCS attraction, we observe this
second contribution, a 
novel pair breaking effect
of the pseudogap. It reflects processes that require a minimal frequency of the
order of $2 \Delta(T)$.
We associate this term with the MIR peak.
Sum rule arguments imply that the larger this MIR peak is,
the smaller the $\omega \approx 0$ contribution becomes;
that is, pseudogap effects lower the dc conductivity
 $\sigma^{dc}$\cite{Ourconduct}.
This transfer of spectral weight can be understood as deriving from the fact
that when non-condensed pairs are present, the number of
fermions available for scattering is decreased; these fermions
are tied up into pairs.

We have derived 
the optical conductivity $\sigma(\omega)$ 
in prevous work
\cite{CSTL05,OurBraggPRL,Kosztin2}.
The
current-current correlation function is
$\tensor{\chi}_{JJ}=\tensor{P}+\frac{\tensor{n}}{m}-C_{\chi}$, where $C_{\chi}$ is associated with 
collective modes, which do not enter above $T_c$ nor in
the transverse gauge below $T_c$.

For notational convenience we define
$E \equiv E_{\mathbf{k}} \equiv \sqrt{\xi_{\textbf{k}}^2 + \Delta^2  }$
as the fermionic excitation spectrum, $\xi_{\textbf{k}}$ is the normal state 
dispersion, $f\equiv f(E)$ is the Fermi distribution function, and
the pairing gap $\Delta^2
= \Delta_{\textrm{sc}}^2 + \Delta_{\textrm{pg}}^2$ is found
\cite{CSTL05,Kosztin2} 
to contain both condensed ($sc$) and non-condensed ($pg$) terms.
In the d-wave case, we write
$\Delta_{\mathbf{k}}=\Delta\varphi_{\mathbf{k}}$,
$\xi_{\mathbf{k}}=-2t(\textrm{cos}k_x+\textrm{cos}k_y)-\mu$,
and $E_{\mathbf{k}}=\sqrt{\xi^2_{\mathbf{k}}+\Delta^2_{\mathbf{k}}}$,
where $\varphi_{\mathbf{k}}=(\textrm{cos}k_x-\textrm{cos}k_y)/2$ is the d-wave
form factor.

The full expression for the current-current response kernel was 
discussed elsewhere \cite{OurBraggPRL,Ourconduct} 
\begin{eqnarray}
\tensor{P}(Q)&\approx&2\sum_K\frac{\partial\xi_{\textbf{k}+\textbf{q}/2}}{\partial\textbf{k}}\frac{\partial\xi_{\textbf{k}+\textbf{q}/2}}{\partial\textbf{k}}\Big[G_KG_{K+Q}\nonumber\\
&+&F_{sc,K}F_{sc,K+Q}
-F_{pg,K}F_{pg,K+Q}\Big]\label{eq:2d}  
\label{eq:2}
\end{eqnarray}
where $Q = (\textbf{q},i\Omega_m)$, $i\Omega_m$ is a bosonic Matsubara frequency, and
the three forms of propagators, introduced in earlier work 
\cite{Ourconduct} are
\begin{eqnarray}
G(K)&=&\Big(i\omega_n-\xi_{\textbf{k}}
+i\gamma-\frac{\Delta_{pg,\textbf{k}}^2}{i\omega_n+\xi_{\textbf{k}}+i\gamma}-\frac{\Delta_{sc,\textbf{k}}
^2}{i\omega_n+\xi_{\textbf{k}}}\Big)^{-1}\nonumber\\
F_{sc}(K) &\equiv&
 -\frac{\Delta_{sc,\textbf{k}}}{i\omega_n+\xi_{\textbf{k}} } \frac{1}{i\omega_n - \xi_{\textbf{k}}
-\frac{\Delta^2_{\textbf{k}}}{i\omega_n+\xi_{\textbf{k}} }} \nonumber\\
F_{pg}(K) &\equiv& - \frac{\Delta_{pg,\textbf{k}}}{i\omega_n+\xi_{\textbf{k}}+ i\gamma}G(K) 
\end{eqnarray}
where $K=(\textbf{k},i\omega_n)$ and $i\omega_n$ is the fermionic Matsubara frequency. The real part of the conductivity can be extracted from $\tensor{P}(Q)$ using the definition $\textrm{Re}\sigma(\omega\neq0)\equiv-\lim_{\textbf{q}\rightarrow0}\textrm{Im}P^{xx}(i\Omega_m\rightarrow\omega+i0^+,\textbf{q})/\omega$.
Here $\gamma$ represents the damping associated principally with the
inter-conversion of fermions and bosons. The first equation representing the full Green's function 
is associated with a BCS self energy ($\propto \Delta_{sc}^2$) and a similar
contribution from the non-condensed pairs ($\propto \Delta_{pg}^2$). The latter
is fairly standard in the literature \cite{Normanphenom} and importantly was derived
microscopically in our earlier work \cite{Maly1}. 
Above, $F_{sc}$ represents the usual Gorkov-like function associated with
condensed pairs and we can interpret $F_{pg}$ as
their non-condensed counterpart.
The full excitation gap
$\Delta(T)$ does not have a strong temperature dependence in the
underdoped regime; below $T_c$ this is because of a conversion
of non-condensed to condensed pairs as $T$ is reduced.

We may rewrite
$\tensor{P}(Q)$
in the regime of very weak dissipation ($\gamma \approx 0$) 
where the
behavior is more physically transparent. For simplicity we will
illustrate this result for $s$-wave pairing
\begin{eqnarray}
& &\tensor{P}(\omega,\mathbf{q})=\sum_{\mathbf{k}}\frac{\mathbf{k}\mathbf{k}}{m^2}\Big[\frac{E_++E_-}{E_+E_-}\big(1-f_+-f_-\big)\nonumber\\                                                                   & &\times\frac{E_+E_--\xi_+\xi_--\delta\Delta^2}{\omega^2-(E_++E_-)^2}-\frac{E_+-E_-}{E_+E_-}\nonumber\\
& &\times\frac{E_+E_-+\xi_+\xi_-+\delta\Delta^2}{\omega^2-(E_+-E_-)^2}\big(f_+-f_-\big)\Big],
\label{eq:5}
\end{eqnarray}
where $f_{\pm}=f(E_{\pm})$ and $\delta\Delta^2=\Delta^2_{\textrm{sc}}-\Delta^2_{\textrm{pg}}$,
$\xi_{\pm}=\xi_{\textbf{k}\pm\textbf{q}/2}$,
and
$E_{\pm}=E_{\mathbf{k}\pm\mathbf{q}/2}$.
Importantly, for this weak dissipation limit, one can analytically show that
\cite{Ourconduct}
the transverse sum rule is precisely satisified. This sum rule
is intimately
connected to the absence above $T_c$ (and presence below) of
a Meissner effect.
The proof depends on
the superfluid density, which at general temperatures is given by
$n_s=(2/3)(\Delta^2_{\textrm{sc}}/m)\sum_{\mathbf{k}}k^2/E^2\Big((1-2f)/2E+\partial f/\partial E\Big)$.
In addition, the total number of particles can be written as
$n=\sum_{\mathbf{k}}\big(1-\xi
(1-2f)/E\big)$.
In this way, it is seen \cite{Ourconduct}
that $\mbox{Re} \sigma(\omega \rightarrow 0)=(\pi n_s/m)\delta(\omega)$.
Since $\Delta_{sc}^2 = \Delta^2 - \Delta_{pg}^2$, one can see that pseudogap effects,
through $\Delta_{pg}^2$, act to lower the superfluid density; the excitation of these
non-condensed pairs provides an additional mechanism, beyond the
fermions, for depleting the condensate with increasing temperature.

We introduce a transport lifetime $\tau=\gamma^{-1}$ into Eq.\ref{eq:5} via the replacement 
$\delta(\omega-(E^+_{\mathbf{k}}\pm E^-_{\mathbf{k}}))=\lim_{\tau\rightarrow\infty}\frac{1}{\pi}\frac{\frac{1}{\tau}}{(\omega-(E^+_{\mathbf{k}}\pm E^-_{\mathbf{k}}))^2+\frac{1}{\tau^2}}$,
to yield (for the more general $d$-wave case) 
\begin{eqnarray}
Re\!~\!\sigma(\!\omega\!\neq\! 0\!)&=& 
\sum_{\mathbf{k}}4\textrm{sin}^2k_xt^2
\Big(\frac{\Delta^2_{\textrm{pg}}(T)\varphi^2_{\mathbf{k}}}{E^2_{\mathbf{k}}}\frac{1-2f(E_{\mathbf{k}})}{2E_{\mathbf{k}}}
\nonumber\\&&\times\big[\frac{\tau}{1+(\omega-2E_{\mathbf{k}})^2\tau^2}
+\frac{\tau}{1+(\omega+2E_{\mathbf{k}})^2\tau^2}\big]
\nonumber\\&&-2\frac{E^2_{\mathbf{k}}-\Delta^2_{\textrm{pg}}\varphi^2_{\mathbf{k}}}{E^2_{\mathbf{k
}}}\frac{\partial
f(E_{\mathbf{k}})}{\partial
E_{\mathbf{k}}}\frac{\tau}{1+\omega^2\tau^2}\Big)
\label{eq:4}
\end{eqnarray}
where we have dropped a small term associated with the derivative of
the d-wave form factor $\varphi^2_{\mathbf{k}}$.
Here  
$\Delta_{sc,\pm}=\Delta_{sc}(T)\varphi_{\textbf{k}\pm\textbf{q}/2}$ and
$\Delta_{pg,\pm}=\Delta_{pg}(T)\varphi_{\textbf{k}\pm\textbf{q}/2}$.
Because of their complexity, we do not include self consistent
impurity effects which, due to bosonic contributions, will require a modification of
earlier work \cite{Lee1993} predicting
$d$-wave fermionic quasi-particles in the ground state. 
Moreover, it seems plausible that non-condensed pairs
may also be associated with these impurity effects, thereby
leading to incomplete condensation and finite $\Delta_{pg}$
in the ground state. 
In general, our calculations
tend to underestimate the very low $T$ spectral weight away from $\omega =0$.

\begin{figure}
\begin{center}
\includegraphics[width=2.7in,clip]
{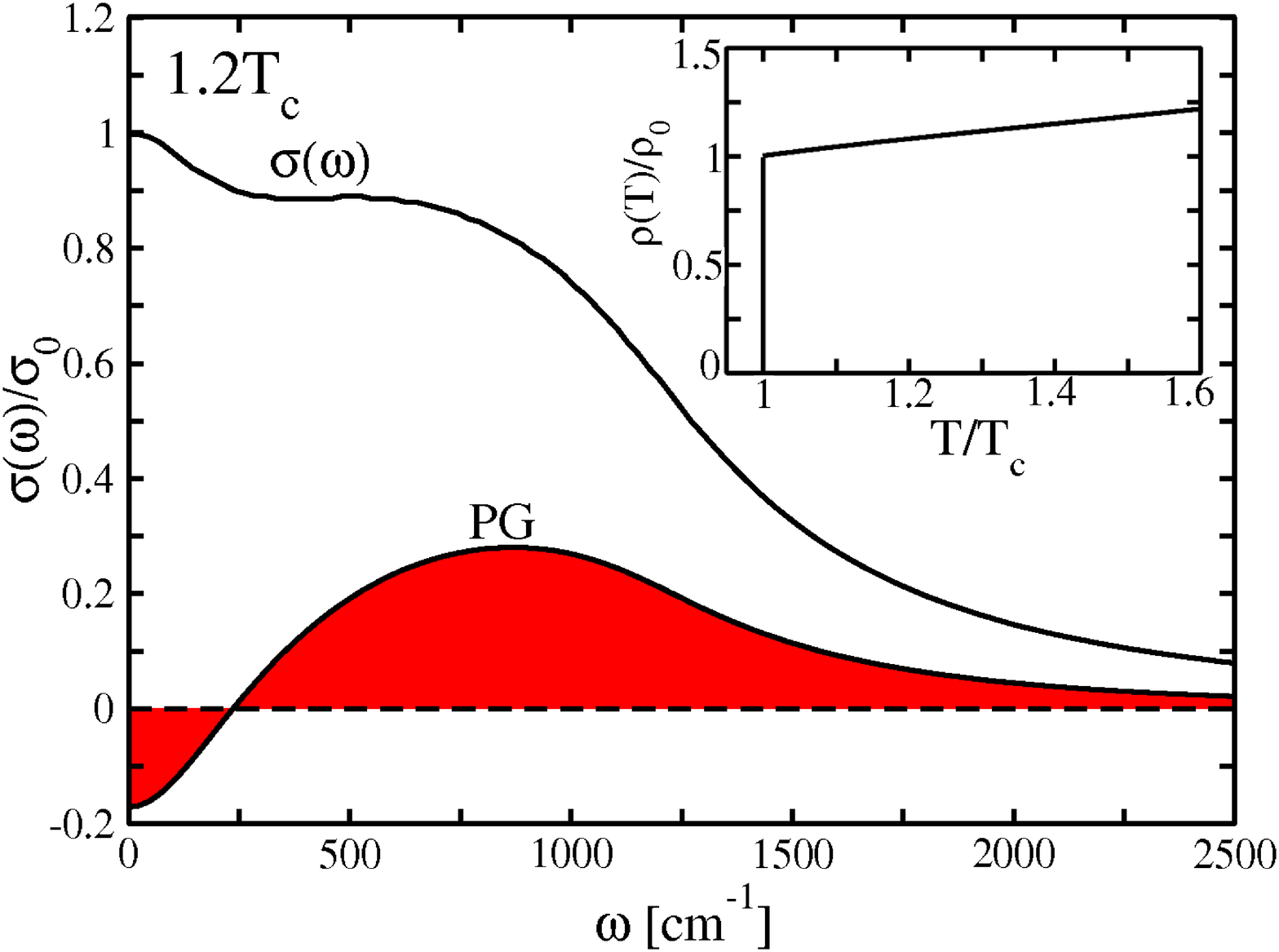}
\includegraphics[width=2.7in,clip]
{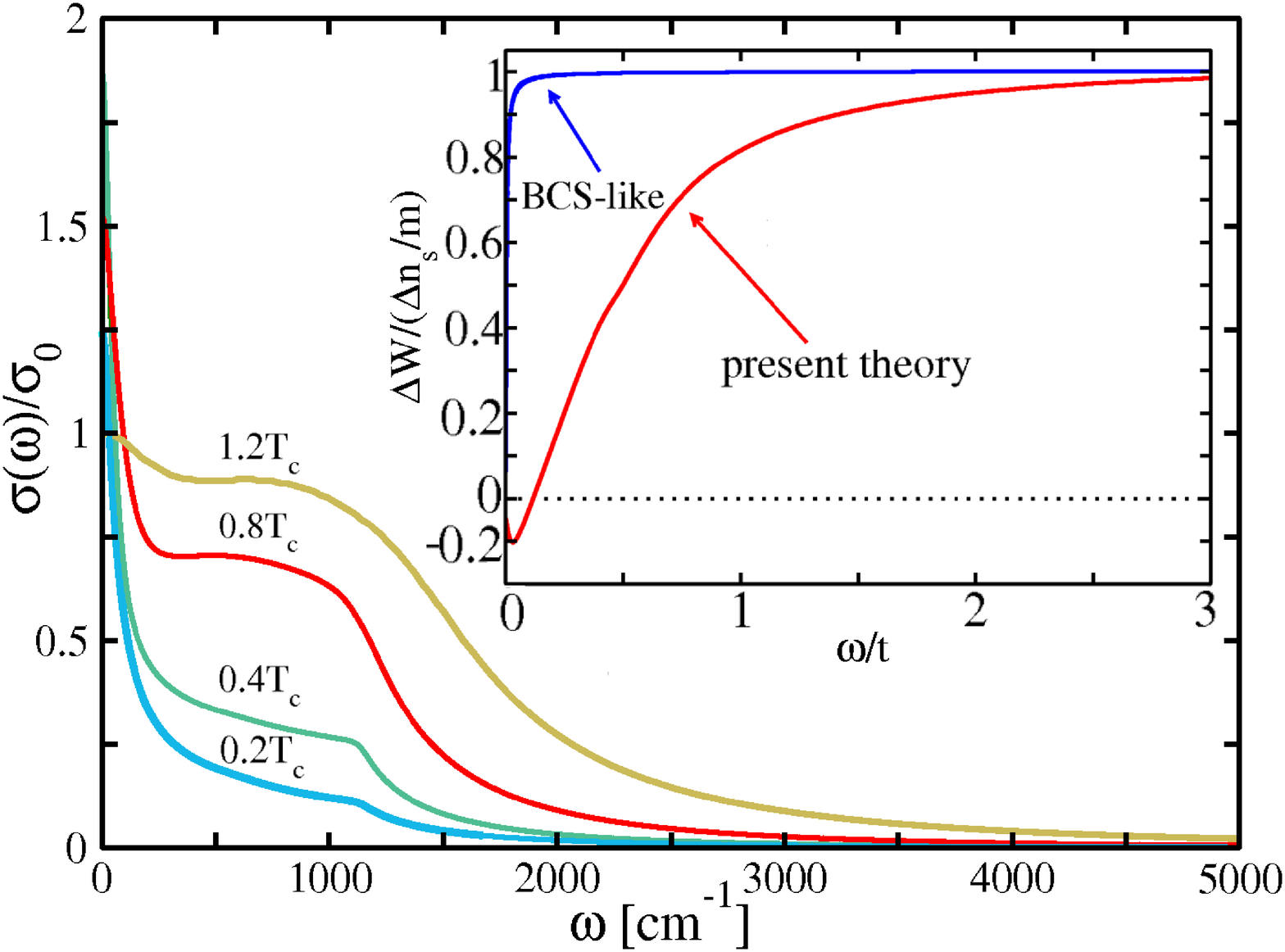}
\caption{Upper panel (top) curve plots $Re~\sigma(\omega)$ for
$T = 1.1T_c$, while the shaded (red) area labelled ``PG" shows
the transfer of spectral weight from low to higher $\omega$ 
associated with non-condensed pairs. 
Inset shows the dc resistivity.
Lower panel plots
$\sigma(\omega)$ at different indicated temperatures.
Normalization
is $\sigma_0=\sigma(0)$ at
$1.2T_c$. The inset shows the difference of spectral weight between $1.4$ and $0.6T_c$ normalized by the difference in superfluid densities. The present theory (red) is contrasted with a BCS-like case (blue) where all explicit $\Delta_{pg}$ contributions are dropped.
}
\label{fig:temp}
\end{center}
\end{figure}

The upper panel in Fig.\ref{fig:temp} 
displays a decomposition of the normal state conductivity vs $\omega$.
The top curve is $Re~\sigma(\omega)$ while the shaded (red) region
labelled ``PG" indicates the contribution from non-condensed
pairs arising from the $F_{pg}$ terms in Eq.~(\ref{eq:2d}).
This figure shows clearly what is implicit in Eq.(\ref{eq:4}), namely that
these pseudogap effects transfer spectral weight from low
to high $\omega$. 
Here the inset plots the resistivity as a function of $T$.

The lower panel in Fig.\ref{fig:temp}
plots the real part
of the optical conductivity versus $\omega$ at the four different temperatures $T/T_c=1.2,0.8,0.4$ and $0.2$. 
There are two peak structures in these plots, the lower Drude-like
peak,  
from the quasi-particle scattering contribution and the upper
peak associated with the breaking of pre-formed pairs. The ``PG''
contribution disappears at the lowest temperatures, as all pairs
go into the condensate. Thus one sees in the figure once
the condensate 
is formed below $T_c$, the low frequency peak 
narrows and increases in magnitude. 
Conversely, the
proportion of the spectral weight residing at high energies on the order of 
$10^{3}cm^{-1}$ increases with temperature. 

To more deeply analyze
this  redistribution of spectral weight, the difference of the
frequency integrated conductivity between $1.4~T_c$ and
$0.6~T_c$ of the
present theory is plotted as a 
function of $\omega/t$ in
the inset of the bottom panel in Figure~\ref{fig:temp}.
Here we define
$W(\omega,T)=(2/\pi)\int_0^{\omega}d\omega'\sigma(\omega',T)$ 
and
$\Delta
W(\omega)=W(\omega,1.4T_c)-W(\omega,0.6T_c)$. 
For comparison, we plot
a counterpart ``BCS-like" spectral weight change which is derived by 
effectively neglecting
the terms involving $\Delta_{pg}^2$ in
Eq.~(\ref{eq:4}).
Both conductivities are normalized by their independently
calculated change in superfluid densities, $\Delta n_s/m$.
The present theory leads to the full integrated
(normalized) spectral weight by $\omega \approx 1~eV$, while the BCS-like
curve counterpart corresponds to $\omega \approx 60~meV$.
One can see that
the presence of non-condensed pairs redistributes an appreciable amount of
spectral weight to higher energies.
Experimentally, there have been claims that very high energy scales
ranging from $1.5-2~eV$ may be needed to satisfy the sum rule.
This figure shows how pseudogap contributions can be, at least
partly, responsible for these high energy scales.

\begin{figure}[htb]
\includegraphics[width=3.5in,clip]
{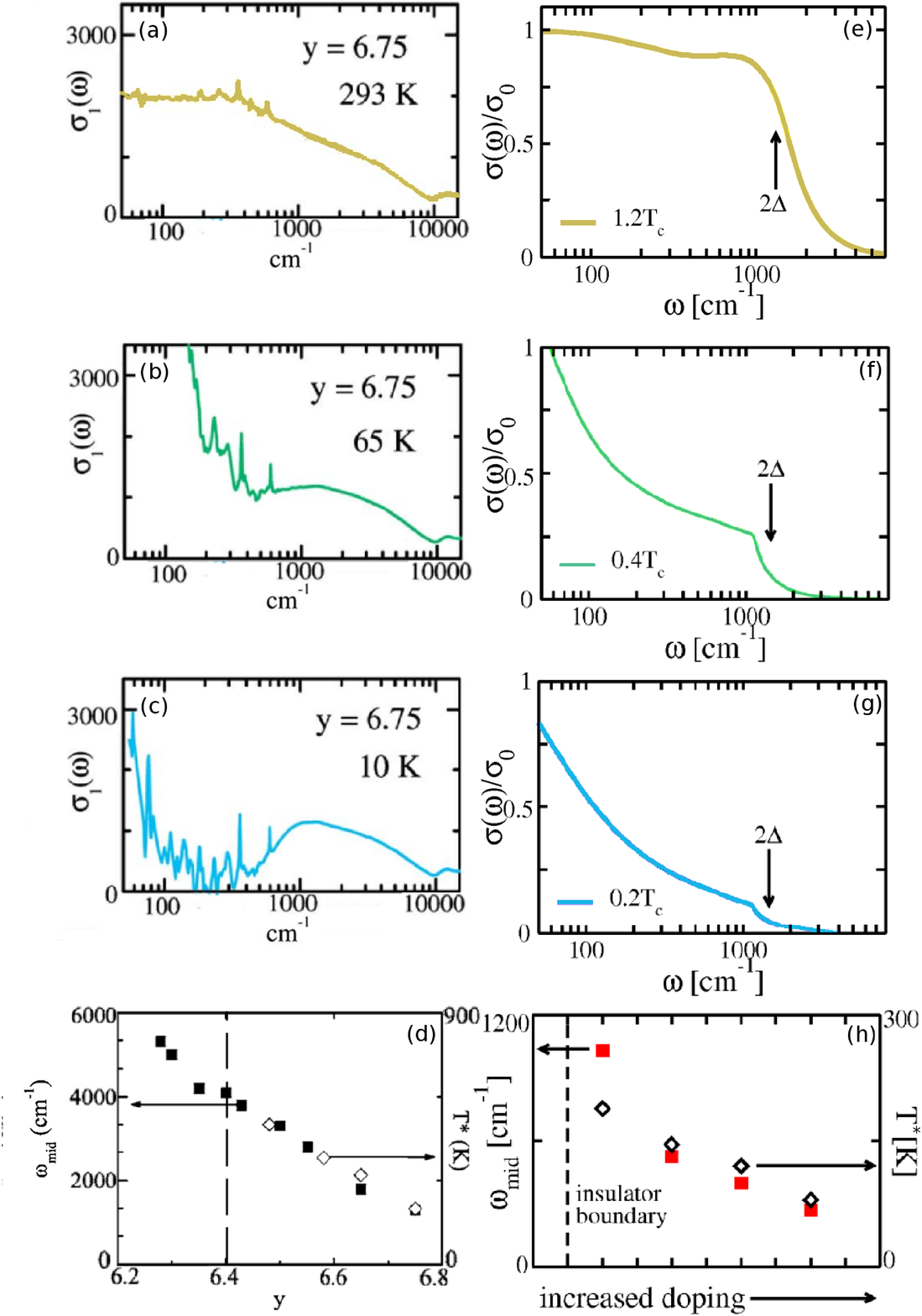}
\caption{The left column consists of figures reproduced from Ref.~\onlinecite{Basov3} and is to be contrasted with the corresponding theoretical results in the right column. Panels (a)-(c) show the optical conductivity for decreasing temperature and (d) plots the MIR peak location $\omega_{mid}$ and $T^*$ as a function of doping. The theoretical results in (e)-(g)  show the optical conductivity for $T/T_c=1.2,0.4,$ and $0.2$ where $\sigma(\omega)$ is normalized by $\sigma_0=\sigma(0)$ at $T=1.2T_c$. The final panel (h) displays $\omega_{mid}$ and $T^*$ as functions of doping. A dashed line in plots (d) and (h) indicates the insulator boundary, which represents the limit of validity for our theory.\label{fig:expthe}}
\label{fig:1}
\end{figure}


We present a more detailed set of comparisons between theory and
experiment in Fig.~\ref{fig:expthe}, where, for the latter, we reproduce the $y=6.75$ plots in Fig. 4 from Ref.~\onlinecite{Basov3} in panels (a)-(c) and the bottom panel of Fig. 5 from the same work in panel (d). Panels (e)-(g) in Fig.~\ref{fig:expthe} are associated with
$T/T_c=1.4,0.4,$ and $0.2$ and should be compared
with the plots in (a)-(c). Here one
sees rather similar trends. Importantly the Drude peak narrows and
increases in height as $T$ decreases. The MIR peak position
is relatively constant, (as seen experimentally)
and in the theory roughly associated with
$2 \Delta$, the value of which is identified in each figure (e)-(g).
That $\Delta(T)$ is roughly constant through the displayed temperature range, 
reflects the inter-conversion of non-condensed to
condensed pairs. 

It should be noted, however, that the height of the MIR peak in the data is more
temperature independent than found in theory. This would seem
to suggest that there are non-condensed pair states at $T=0$ perhaps
associated with inhomogeneity or localization \cite{Lee1993} effects.
This interpretation of the optical
data appears consistent with our previous studies \cite{ourarpes} of angle resolved
photoemission (ARPES) data from which we have inferred
that the ground state
in strongly underdoped samples may not be the fully condensed d-wave
BCS phase. Rather there may be some non-condensed pair
or pseudogap effects which persist to T=0. In ARPES experiments
one could attribute this persistence to 
the fact that the $T=0$ gap shape is distorted relative to
the more ideal $d$-wave form found in moderately underdoped systems
\cite{ShenNature}. Similar observations are made from STM
experiments \cite{Yazdani3}.

We show in Fig.~\ref{fig:expthe}(h) a plot of the MIR peak location $\omega_{mid}$ as a function of $T^*$ as calculated in our theory; this
plot suggests that the
MIR peak position scales (nearly linearly) with the pairing gap or equivalently
with $T^*$. This observation is qualitatively similar
(within factors of 2 or 3) to Fig.\ref{fig:expthe}(d), reproduced from
Ref.\onlinecite{Basov3}.
Finally, we stress that we have investigated the effects of varying $\gamma$
as well as its $T$ dependence and find that
our results in
Fig.~\ref{fig:expthe}
remain very robust.


At the core of interest in the optical conductivity is
what one can learn about the origin of the pseudogap.
We earlier discussed problematic aspects of
alternative scenarios for the two component
optical response.
We reiterate that the observed tight correlation with the two component 
optical response 
and the presence of a pseudogap \cite{Basov3}
is natural in the present theory, where the MIR
peak is to be associated
directly with the breaking
of meta-stable pairs. Such a contribution
does not disappear below $T_c$, until all pairs are condensed.
In summary, our paper appears compatible with the
very important experimental conclusion in Ref.~\onlinecite{Basov3} that
``Our findings suggest that any explanation [of the MIR peak] 
should take into account the correlation betwen the formation of the mid IR 
absorption and the development of the pseudogap."

This work is supported by NSF-MRSEC Grant
0820054 and we thank Vivek Mishra for helpful
conversations. 
C.C.C. acknowledges the support of the U.S.
Department of Energy through the LANL/LDRD Program.

\appendix

\section{Appendix A: Further numerical studies}

\begin{widetext}
In this Appendix we present a comparison figure, Fig.\ref{fig:3}, for the optical conductivity
in the case where we take the parameter $\gamma$ to be constant in temperature for three different values of $\gamma$. The first column in Fig.\ref{fig:3} reproduces the experimental data from Ref.~\onlinecite{Basov3}.The remaining columns show the theoretical results for decreasing values of $\gamma$. Each row corresponds to decreasing temperature from top to bottom.  
This shows that are results are robust over a large range of $\gamma$ and very much
independent of what values or temperature dependences are
assumed for the lifetime broadening. 

\begin{figure}
\begin{center}
\includegraphics[width=6.0in,clip]
{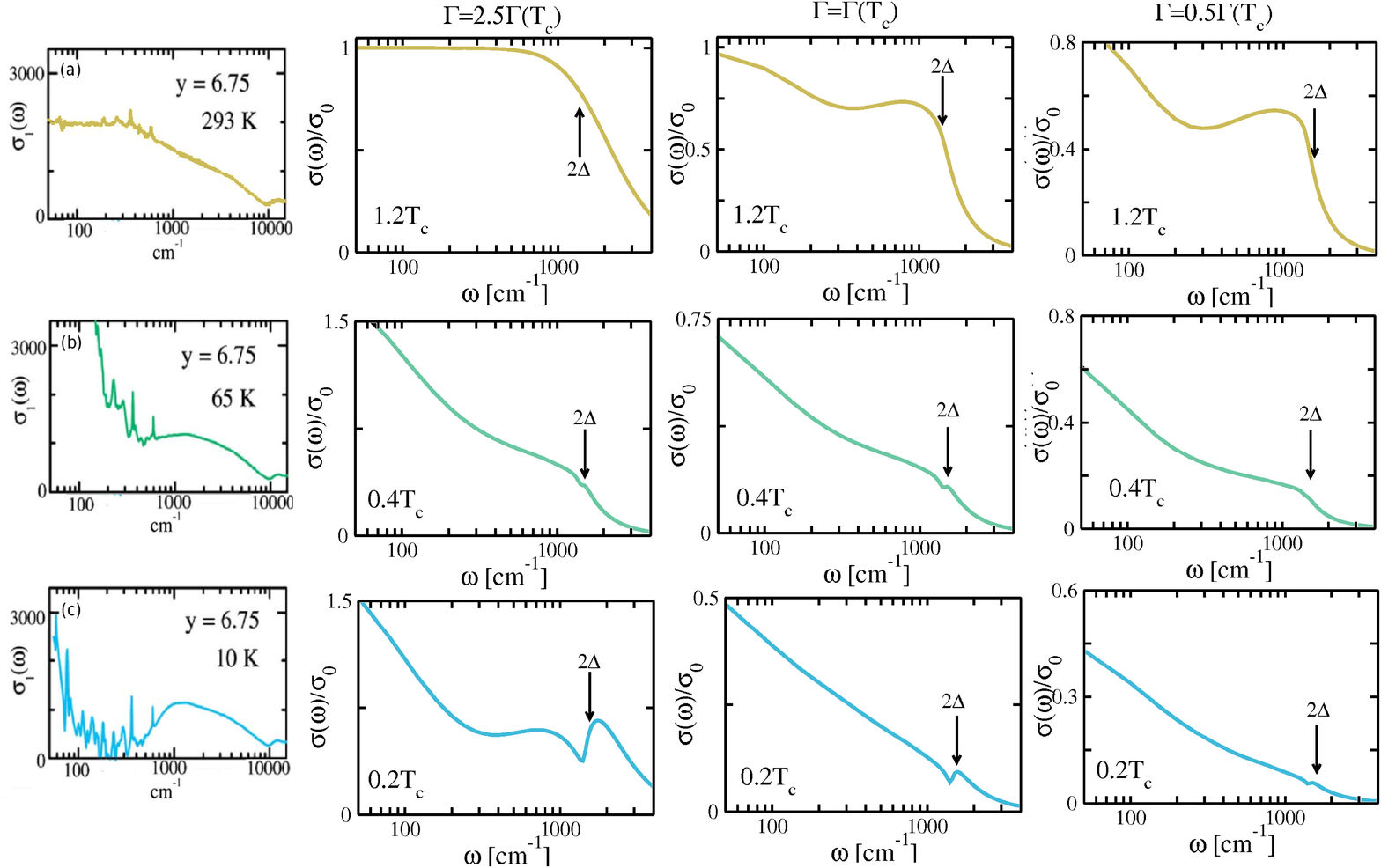}
\caption{Illustrative figure for the case of constant gamma. The numerical values indicated
are quoted relative to those we showed in the paper.\label{fig:3}}
\label{fig:7}
\end{center}
\end{figure}

\end{widetext}

\bibliography{Literatureopt.bib} 

\end{document}